\documentclass{iopart}
\usepackage{iopams,amsmath,bm,algorithm,algorithmic}
\usepackage{array,eqparbox,dcolumn,multirow,slashbox}
\usepackage[pdftex]{graphicx,color}
\graphicspath{{figures/pdf/},{figures/jpg/}}
\DeclareGraphicsExtensions{.pdf}
\usepackage{subfig}
\newtheorem{theorem}{Theorem}
\def\vec#1{{\bf#1}}

\def\op#1{\hat{#1}}

\def\ket#1{| #1 \rangle}
\def\bra#1{\langle #1 |}

\def\ave#1{\langle #1 \rangle}

\def\Tr{\operatorname{Tr}}

\def\diag{\operatorname{diag}}

\font\liouvop=eufm10

\def\D{{\text{\liouvop D}}}

\def\RR{\mathbb{R}}

\def\LL{\mathfrak{L}}

\def\LL{\mathbf{L}}
\def\DD{\mathbf{D}}
\def\H{\mathcal{H}}

\def\sx{\op{\sigma}_x}
\def\sy{\op{\sigma}_y}
\def\sz{\op{\sigma}_z}

\def\Id{\mathbb{I}}

\def\ss{\rm ss}

\begin{document}

\title{Open quantum system identification}

\author{Sophie G. Schirmer} 
\address{Dept of Physics, College of Science, Swansea University,
         Singleton Park, Swansea, SA2 8PP, United Kingdom}
\eads{\mailto{sgs29@swan.ac.uk}}

\author{Daniel K. L. Oi}
\address{SUPA Department of Physics, University of Strathclyde,
  Glasgow G4 0NG, United Kingdom}

\author{Weiwei Zhou, Erling Gong, Ming Zhang}
\address{College of Mechatronic Engineering and Automation, 
         National University of Defense Technology, Changsha 410073, China}

\date{\today}

\begin{abstract}
  Engineering quantum systems offers great opportunities both
  technologically and scientifically for communication, computation,
  and simulation. The construction and operation of large scale
  quantum information devices presents a grand challenge and a major
  issue is the effective control of coherent dynamics. This is often
  in the presence of decoherence which further complicates the task of
  determining the behaviour of the system. Here, we show how to
  determine open system Markovian dynamics of a quantum system with
  restricted initialisation and partial output state information.
\end{abstract}


\submitto{\NJP}

\maketitle

\section{Introduction}

Recently, much effort has been put into the design and realization of
large scale quantum devices operating in the coherent regime. This has
been spurred by the possibilities offered by quantum communication and
information processing, from secure transmission, simulation of quantum
dynamics, and the solution of currently intractable mathematical
problems~\cite{NatureReview2008}.  Many different physical systems have
been proposed as basic architectures upon which to construct quantum
devices, ranging from atoms, ions, photons, quantum dots and
superconductors. For large scale commercial applications, it is likely
that this will involve scalable engineered and constructed devices with
tailored dynamics requiring precision control.

Due to inevitable manufacturing tolerances and variation, each device
will display different behaviour even though they may be nominally
``identical''. For their operation, they will need to be characterised
as to their basic properties, response to control fields, and noise or
decoherence~\cite{BIRS2007}. We may also need to know how ideal is the
system in the first place, for instance the effective Hilbert space;
what we may assume to be a qubit may have dynamics involving more than
two effective levels~\cite{GoodQubit}. Extracting this information
efficiently and robustly is crucial.

In a laboratory setting, an experimentalist may have access to many
tools with which to study a system, e.g. spectroscopy and external
probes.  In a production setting, provision of these extra resources
may be difficult, expensive, or impossible to integrate with the
device. It is therefore an important to understand what sort of
characterisation can be performed simply using what is available
\emph{in situ}. Ideally, we would also like to be able to characterise
the performance of a device with as little prior parameter data,
e.g. how it responds to control fields, if this is the information
which we are trying to obtain in the first place. Characterisation
using only the \emph{in situ} resources of state preparation and
measurement, even where it is possible, is challenging, due to the
increasing complexity of the signals, number of parameters to be
estimated and the complexity of reconstructing a valid Hamiltonian
from the resulting signal parameters.  Robust and efficient methods of
data gathering and analysis, preferably in as an automated way as
possible, are therefore essential.

System identification attempts to extract dynamics from minimal
pre-existing system control resources. There is a trade-off between the
universality of the dynamics that can be identified versus the
resources available in terms of initialisation, access, and
measurement. Here, we explore the issue of what can be extracted in
practice, leaving aside the question of efficiency for future work. We
show how we can extract the dynamics of an open quantum system with
restricted state initialisation and only partial information of the
evolving state. This relies on prior knowledge of the structure of the
decoherence processes.  Our results show that in some cases full state
information without prior knowledge can lead to worse estimation than
partial state information combined with limited a-priori knowledge about
the expected model type.

\subsection{Quantum Process Tomography and System Identification}

In quantum process tomography (QPT)~\cite{QPT}, what is determined is
the discrete map between initial and final states. This requires in
general the preparation of a complete set of input density operators,
and quantum state tomography (QST)~\cite{QST} of each of the output
states, the result of which gives a completely positive map.  However,
this does not fully characterise the dynamical behaviour of the system
such as the Hamiltonian or Lindblad operators for Markovian systems.  In
principle, by conducting QPT at various times, it is possible to
construct a time dependent set of Kraus operators~\cite{Havel}, hence
capturing the most general (not necessarily Markovian) open systems
evolution but this is costly in both terms of initial state preparation
and QST.  We therefore look for methods which require fewer resources to
determine the dynamics of a system, though perhaps at the expense of the
ability to handle the most general evolutions, considering that for many
systems the physics is sufficiently known so that the structure of the
dynamics can be constrained, even if the precise values (which we are
trying to determine) may not be known.

Previous work has examined several system identification scenarios.
Hamiltonian identification for a single qubit was introduced
in~\cite{HamilIdent1,HamilIdent2,HamilIdent3}, Hamiltonian
identification in the presence of decoherence was examined
in~\cite{HamilDecoh}, extension to higher dimensional systems and the
use of Bayesian/Maximum Likelihood techniques in~\cite{TwoQubits}
and~\cite{QutritDecoh}, and model-based system parameter reconstruction
in~\cite{Kavli}. Other work has also studied identification of system
Hamiltonians for spin networks with restricted access~\cite{Burgarth1, 
Burgarth2} and limited system initialisation~\cite{Paternostro}.

In the setting where preparation and measurement are limited to a single
fixed basis it was found that signal and parameter complexity proved to
be a major challenge even for relatively small systems and Hamiltonian
evolution~\cite{TwoQubits}.  Including open system behaviour magnifies
the problem even further. Besides increasing the number of parameters to
be estimated, damping reduces the signal length and limits the amount of
information that can be extracted~\cite{HamilDecoh}. To make the problem
tractable we can try to incorporate prior information about the
structure of the dynamics~\cite{QutritLeak1,QutritLeak2}.  In this case
we find that even when only partial projective measurement data is
available the signal may contain sufficient information to enable
Hamiltonian reconstruction~\cite{Kavli}.

Here we examine problem of determining open quantum system dynamics
based on restricted initial state preparation and incomplete state
reconstruction, e.g., limited to measurements of a single observable.
Using a combination of Bayesian/Maximum likelihood estimation of signal
parameters~\cite{Bretthorst} coupled with analysis of the equations of
motion in the Laplace domain we study the problem of identifiability of
models, comparing scenarios involving different types and amounts of
information available for reconstruction.  The results are applied to 
the problem of identifying the dynamics of a qubit in a Markovian
environment to reconstruct both Hamiltonian and decoherent processes
in various settings.

\section{Master equation for Markovian Case}

The evolution of an $N$-level open quantum system in a Markovian
environment is given by the dissipative master equation
\begin{equation}
\label{eq:LME} \frac{d}{dt}\rho(t) 
 = -\frac{i}{\hbar}[H,\rho] + \sum_{n,m=1}^{N^2-1} f_{mn} \D(F_m,F_n) \rho(t)
\end{equation}
where $\rho(t)$ is the state of the system at time $t$ and $H$ is a
Hermitian matrix representing the effective Hamiltonian.  In the
following we shall choose units such that $\hbar=1$ and drop $\hbar$.
The requirement of completely positive evolution necessitates that the
super-operators $\D(F_m,F_n)\rho(t)$ be of the form~\cite{Lindblad,GFKS,BP}
\begin{equation}
\label{eq:GKS}
  \D(F_m,F_n)\rho(t) = F_n\rho F_m^\dag 
 -\tfrac{1}{2} (\rho F_m^\dag F_n + F_m^\dag F_n\rho)
\end{equation}
where $\{F_n\}_{n=1}^{N^2-1}$ is an orthonormal basis for all trace-zero
operators on $\H$, and the coefficient matrix $(f_{mn})$ is positive.
The latter requirement implies $f_{mn}^*=f_{nm}$, i.e., we can write
$f_{mn}=a_{mn}+ib_{mn}$, where $(a_{mn})$ is a symmetric real matrix and
$(b_{mn})$ is a real-anti-symmetric matrix.

Taking $\{\sigma_k\}_{k=1}^{N^2}$ to be a basis for the Hermitian
matrices on the Hilbert space $\H$ the master equation~(\ref{eq:LME})
with dissipation term~(\ref{eq:GKS}) can be written in coordinate form
as a linear matrix differential equation (DE)~\cite{Lendi}
\begin{equation}
 \tfrac{d}{dt}\vec{\tilde{r}} = (\LL+\DD)\vec{\tilde{r}},
\end{equation}
where $\vec{\tilde{r}}=(r_n)\in\RR^{N^2}$ with $r_n=\Tr(\sigma_n\rho)$
and $\LL$ and $\DD$ are $N^2\times N^2$ (real) matrices with entries
\begin{subequations}
\begin{align}
\label{eqn:LD}
  L_{mn} &= \Tr(i H [\sigma_m,\sigma_n]) 
          = \sum_{n'} h_{n'} \Tr(i \sigma_{n'} [\sigma_m,\sigma_n] ), \\
  D_{mn} &= \sum_{m',n'} f_{m'n'} \Tr(F_{m'}^\dag \sigma_m F_{n'} \sigma_n
                  -\tfrac{1}{2} F_{m'}^\dag F_{n'} \{\sigma_m,\sigma_n\}).
\end{align}
\end{subequations}
Here $\{A,B\}=AB+BA$ and $[A,B]=AB-BA$ are the usual matrix
anti-commutator and commutator, respectively.  If we choose a basis such
that $\sigma_{N^2}=\frac{1}{\sqrt{N}}\Id$ and the remaining basis
elements form a basis for the trace-zero Hermitian matrices then
\begin{equation}
  r_{N^2} = \tfrac{1}{\sqrt{N}}\Tr(\rho)= \tfrac{1}{\sqrt{N}}
\end{equation}
is constant, i.e., $\dot{r}_{N^2}=0$, and we can define a reduced
so-called Bloch vector $\vec{r}=(r_1,\ldots,r_{N^2-1})^T$~\cite{Kimura},
and rewrite the linear matrix DE for $\vec{\tilde{r}}$ as affine-linear
DE
\begin{equation}
  \label{eq:Bloch}
  \dot{\vec{r}}(t) = A \vec{r}(t) + \vec{c},
\end{equation} 
where $A$ is an $(N^2-1)\times (N^2-1)$ real matrix with $A_{mn}=L_{mn}+
D_{mn}$ and $c_m=\frac{1}{\sqrt{N}} D_{m N^2}$.  We can without loss of
generality choose $F_{m}=\sigma_n$.

The objective of complete identification of the dynamics is thus reduced
to identifying the coefficients $h_{n'}$ and $f_{m'n'}$ of the
$N^2\times N^2$ (Hermitian) coefficient matrix.  Positivity constraints
restrict the parameter space further to a convex subset of $\RR^{N^4}$.

\section{Identifiability given Complete State Information}
\label{sec2}

The Bloch equation for dissipative quantum systems always has at least
one steady state $\vec{r}_{\ss}$~\cite{SteadyState} for which we have
\begin{equation}
 \vec{0} = \dot{\vec{r}}_{\ss} = A \vec{r}_{\ss} + \vec{c}.
\end{equation}
This implies $\vec{c}=-A \vec{r}_{\ss}$ and shows that the Bloch vector 
$\vec{r}(t)$ is given by
\begin{equation}
  \label{eq:r(t)-general}
  \vec{r}(t) = \vec{r}_{\ss} + e^{tA} (\vec{r}(0)-\vec{r}_{\ss}).
\end{equation}
We note that if there are two steady states $\vec{r}_{\ss}^{(1,2)}$ then
$A \vec{\Delta r}_{\ss} = A (\vec{r}_{\ss}^{(1)}-\vec{r}_{\ss}^{(2)})=0$
shows that $A \vec{r}_{\ss}^{(1)} = A \vec{r}_{\ss}^{(2)}$ and thus
$\vec{c}$ is well-defined and
\begin{align*}
  \vec{r}(t) 
  &= \vec{r}_{\ss}^{(2)} + e^{tA} (\vec{r}(0)-\vec{r}_{\ss}^{(2)})\\
  &= \vec{r}_{\ss}^{(1)} + \vec{\Delta r}_{\ss}
     + e^{tA} (\vec{r}(0)-\vec{r}_{\ss}^{(1)} - \vec{\Delta r}_{\ss})\\
  &= \vec{r}_{\ss}^{(1)} + e^{tA} (\vec{r}(0) - \vec{r}_{\ss}^{(1)})
\end{align*}
noting that $A\vec{x}=0$ implies $\exp(tA)\vec{x}=1\vec{x}$.


Generically the superoperator $A$ has $d=N^2-1$ distinct eigenvalues
$\lambda_n$ with corresponding eigenvectors $\vec{v}_n$.  The
eigenvalues and eigenvectors may be complex but as $A$ itself is real,
complex eigenvalues must occur in complex conjugate pairs.  Suppose
there are $p<d/2$ pairs of complex eigenvalues $\lambda_n^{\pm} =
\gamma_n\pm i\omega_n$ and $d-2p$ real eigenvalues $\lambda_n=\gamma_n$
for $n=p+1,\ldots,d-p$.  Writing the corresponding eigenvectors as
$\vec{v}_n^{\pm} =\vec{x}_n\pm i\vec{y}_n$ for $n\in I_1=\{1,\ldots,p\}$
and $\vec{v}_n= \vec{x}_n$ for $n \in I_2=\{p+1,\ldots,d-p\}$ with
$\vec{x}_n,\vec{y}_n\in\RR^d$, and noting that $A^*=A$ shows that
\begin{align*}
  A (\vec{x}+i\vec{y}) = (\gamma+i\omega) (\vec{x}+i\vec{y}) &
\quad \Rightarrow \quad 
  A (\vec{x}-i\vec{y}) = (\gamma-i\omega) (\vec{x}-i\vec{y}).
\end{align*}
The vectors $\{\vec{x}_n\}_{n=1}^{d-p}$ and $\{\vec{y}_n\}_{n=1}^p$
together form a (non-orthogonal) complete basis for $\RR^d$ and we can
expand the initial state $\vec{r}_0$ and steady state $\vec{r}_{\ss}$
with respect to it
\begin{align}
  \vec{r}_{\ss} = \sum_{n=1}^{d-p} \beta_n\vec{x}_n
                   +\sum_{n=1}^{p} \beta_n' \vec{y}_n, &&
  \vec{r}_0     = \sum_{n=1}^{d-p} (\alpha_n+\beta_n)\vec{x}_n
                   +\sum_{n=1}^{p} (\alpha_n'+\beta_n') \vec{y}_n,
\end{align}
where $\alpha_n$ and $\beta_n$ are real coefficients.  This yields
\begin{equation}
\begin{split}
\label{eq:ro-rss}
 \vec{\Delta}(t) = \vec{r}_0 -\vec{r}_{\ss}
 &= \sum_{n\in I_1} \alpha_n \vec{x}_n + \alpha_n' \vec{y}_n +
 \sum_{n\in I_2} \alpha_n \vec{x}_n\nonumber\\
 &= \sum_{n\in I_1} \frac{\alpha_n}{2} [\vec{v}_n^++\vec{v}_n^-] 
                  + \frac{\alpha_n'}{2i} [\vec{v}_n^+-\vec{v}_n^-]
    + \sum_{n\in I_2} \alpha_n \vec{v}_n \nonumber\\
 &= \sum_{n\in I_1} \tfrac{1}{2}(\alpha_n-i\alpha_n') \vec{v}_n^+
                   +\tfrac{1}{2}(\alpha_n+i\alpha_n') \vec{v}_n^-
   +\sum_{n\in I_2} \alpha_n \vec{v}_n\nonumber\\
 &= \sum_{n\in I_1} \xi_n \left( \vec{\tilde{v}}_n^+ + \vec{\tilde{v}}_n^- \right)
    + \sum_{n\in I_2} \xi_n \vec{\tilde{v}}_n
\end{split}
\end{equation}
where $\vec{\tilde{v}_n}^{\pm}=\tfrac{1}{2}(\alpha_n\mp i\alpha_n')
\vec{v}_n^{\pm}$ and $\vec{\tilde{v}_n}=\alpha_n\vec{v}_n$ and $\xi_n=0$
if $\alpha_n=\alpha_n'=0$ and $\xi_n=1$ otherwise.  Inserting the last
expression into (\ref{eq:r(t)-general}) and recalling $A \vec{v}_n^{\pm}
= (\gamma_n\pm i\omega_n)\vec{v}_n^\pm$ we obtain
\begin{equation}
 \label{eq:r(t)}
 \begin{split} 
  \vec{r}(t) =& \vec{a}_0 + \sum_{n=1}^p \xi_n e^{\gamma_n t} 
               \left[\vec{a}_n \cos(\omega_n t) + \vec{b}_n \sin(\omega_n t)\right]
              + \sum_{n=p+1}^{d-p} \xi_n \vec{a}_{n} e^{\gamma_n t}
\end{split}
\end{equation}
where the coefficient vectors are $\vec{a}_0 = \vec{r}_{\ss}$,
$\vec{a}_n=2\vec{\tilde{x}_n}$, $\vec{b}_n=-2\vec{\tilde{y}_n}$ for
$n\in I_1$, $\vec{a}_n=\vec{\tilde{v}}_n$ for $n\in I_2$, and
$\vec{\tilde{v}_n^{\pm}}= \vec{\tilde{x}}_n \pm i \vec{\tilde{y}}_n$
with $\vec{\tilde{x}}_n = \tfrac{1}{2}(\alpha_n
\vec{x}_n+\alpha_n'\vec{y}_n)$, $\vec{\tilde{y}}_n =
\tfrac{1}{2}(-\alpha_n' \vec{x}_n+\alpha_n\vec{y}_n)$.  The coefficient
vectors $\vec{a}_0$, $\vec{a}_n$ and $\vec{b}_n$ can be estimated along
with the parameters $\omega_n$ and $\gamma_n$ using Bayesian estimation,
choosing the basis functions 
\begin{subequations}
\begin{align}
   g_0         &= 1                                 \\
   g_{2n-1}(t) &= e^{\gamma_n t} \sin(\omega_n t) &&\quad n=1,\ldots, p \\
   g_{2n}(t)   &= e^{\gamma_n t} \cos(\omega_n t) &&\quad n=1,\ldots, p \\
   g_{2p+n}(t) &= e^{\gamma_n't}, && \quad n=1,\ldots,N-1. 
\end{align}
\end{subequations}
Thus, provided $\xi_n\neq 0$ for all $n$, we can determine all of the
eigenvectors $\vec{\tilde{v}}_n^{\pm}$ and eigenvalues of $A$, so that
$A=S D S^{-1}$ with
$S=(\vec{\tilde{v}}_1^+,\vec{\tilde{v}}_1^-,\ldots\vec{\tilde{v}}_p^+,
\vec{\tilde{v}}_p^-,\vec{\tilde{v}}_{p+1},\ldots,\vec{\tilde{v}}_{p+d})$
being a matrix whose columns are the complex eigenvalues of $A$ and
$D=\diag(\lambda_1^+,\lambda_1^-, \ldots \lambda_p^+,\lambda_p^-,
\gamma_{p+1},\ldots,\gamma_{p+d})$ is a diagonal matrix with the
corresponding eigenvalues $\lambda_n^\pm =\gamma_n\pm i\omega_n$.  $S$
will be invertible provided $\xi_n\neq 0$ for all $n$ so that we have
determined all eigenvectors.

\begin{theorem}
The trajectory $\vec{r}(t)$ of an initial state uniquely determines the
$A$ and $\vec{c}=-A\vec{a}_0$ provided $\vec{r}(0)-\vec{r}_{\ss}$, where
$\vec{r}_{\ss}$ is a steady state, has non-zero overlap with all 
eigenvectors of $A$.
\end{theorem}


If $A$ has fewer than $N^2-1$ distinct eigenvalues then
$\vec{\Delta}(t)=\vec{r}(t)-\vec{r}_{\ss}$ still satisfies a homogeneous
linear equation $\dot{\vec{\Delta}}(t)=A\vec{\Delta}(t)$ and we have
$A=SJS^{-1}$, where $S$ is a matrix determined by the (generalised)
eigenvectors but $J=\diag(J_n)$ is the Jordan normal form of $A$
consisting of irreducible Jordan blocks $J_n$ of dimension $k_n$ with
some eigenvalue $\lambda_n$.  In this case we have $\vec{\Delta}(t)=S
e^{Jt}S^{-1} \vec{\Delta}(0)$, where $S$ is a matrix whose columns are
generalised eigenvectors $\vec{v}_{n,k}^{\pm}$ satisfying
$(A-\lambda_n^\pm I)^k\vec{v}_{n,k}^{\pm}=0$, $(A-\lambda_n^\pm
I)^{k-1}\vec{v}_{n,k}^{\pm}\neq 0$.  The generalised eigenvectors can be
chosen such as to be span $\RR^{N^2-1}$. The matrix exponential $e^{Jt}$
is block-diagonal with $k_n$ dimensional blocks
\begin{equation}
  \label{eq:En}
  E_n(t) = e^{t\gamma_n}
  \begin{bmatrix} R_n & t R_n & \frac{1}{2} t^2 R_n
                    & \frac{1}{6} t^3 R_n &\ldots \\
                  0 & R_n & tR_n & \frac{1}{2} t^2R_n &\ldots\\
                  0 & 0 & R_n & tR_n & \ldots \\
                  \vdots & \vdots & \vdots & \ddots
 \end{bmatrix}, 
\end{equation}
where 
\begin{equation}
 R_n = 1 \textrm{ for } \omega_n=0, \qquad 
 R_n = \begin{pmatrix}
            \cos(\omega_n t) & -\sin(\omega_n t) \\
            \sin(\omega_n t) & \cos(\omega_n t)
	  \end{pmatrix}
 \textrm{ for } \omega_n\neq0.
\end{equation}

For non-generic $A$ there are many different subcases that arise even in
the simplest case of a single qubit.  These are discussed in detail in
\ref{App:non-generic}.  Most of these special cases rarely arise in
practice and the nature of the positivity constraints impose further
restrictions.  Physically allowed examples of non-diagonalisable Bloch
matrices do exist.  For qubits they involve extremely high decoherence
rates leading to over-damped dynamics~\cite{Lendi}, which are usually not
of interest for quantum information.  In higher dimensions there are
more possibilities for $A$ to non-diagonalisable although generically
$A$ is still diagonalisable.  The analysis in the appendix suggests that
the trajectory $\vec{r}(t)$ of one initial state still uniquely
determines $A$ and $\vec{c}=-A\vec{a}_0$ even in the non-generic case
provided (i) $A$ does not have two or more generalised eigenvectors of
the same degree $k$ corresponding to the same eigenvalue $\lambda$, and
(ii) $\vec{r}(0)-\vec{r}_{\ss}$, where $\vec{r}_{\ss}$ is a steady
state, has non-zero overlap with all (generalised) eigenvectors of $A$.

\section{Identifiability given Partial State Information}

A more interesting question is how much information about the dynamical
generators we can obtain from partial state information such as
measurement of a single observable.  To address this problem it is
useful to consider the identification problem in the Laplace
domain~\cite{HamilDecoh}.  The \emph{Laplace Transform} of the Bloch
equation~(\ref{eq:Bloch}) gives
\begin{equation}
  \label{eq:BlochL}
  s^2 \vec{R}(s) - s\vec{r}(0) = A \vec{R}(s) s + \vec{c}
                               = T \begin{pmatrix} s\vec{R}(s) \\ 1\end{pmatrix}
\end{equation}
where $\vec{R}(s)$ is the Laplace transform of $\vec{r}(t)$ and
$T=(A,c)$ is the matrix $A$ with the column vector $\vec{c}$ appended.
Thus, given $\vec{R}(s)$ and $\vec{r}(0)$ then $T$ and thus $A$ and
$\vec{c}$ are in principle fully determined.  Indeed we only need to
know $\vec{R}(s)$ for $d+1$ values of $s_k$ in general.  Define the
$d+1\times d+1$ matrices
\begin{subequations}
  \begin{align}
  C &= \begin{pmatrix}
	s_1 \vec{R}(s_1)&\ldots &
	s_{d+1} \vec{R}(s_{d+1})\\  & \ldots &1\end{pmatrix}, \\
  D &= \begin{pmatrix} s_1^2\vec{R}(s_1)-s_1\vec{r}(0),
	&\ldots,s_{d+1}^2\vec{R}(s_{d+1})-s_{d+1}\vec{r}(0)\end{pmatrix}.
\end{align}
\end{subequations}
We have $D=TC$ and if $\det(C)\neq 0$ then $T=D C^{-1}$ and thus we have
fully determined both $A$ and $\vec{c}$.  Thus if we could measure in
the Laplace domain rather than the time domain only $d+1$ measurements
would be required to completely determine the dynamics although it not
easy to see how to perform such measurements in practice.

%

But (\ref{eq:BlochL}) can also be useful if we have incomplete
information about the state, e.g., if we can only measure some
components of the Bloch vector.  We can use (\ref{eq:BlochL}) to express
the unknown components in terms of the known ones.  For example, suppose
we know only the last component $R_d(s)$.  Let $\vec{R}'(s)$,
$\vec{r}'(0)$ and $\vec{c}'$ denote the first $d-1$ components of
$\vec{R}(s)$, $\vec{r}(0)$ and $\vec{c}$, and partition the matrix $A$
as follows
\begin{equation}
  A = \begin{pmatrix} A' & \vec{a} \\ \vec{b}^T & a_{dd} \end{pmatrix}.
\end{equation}
Then (\ref{eq:BlochL}) can be rewritten as
\begin{subequations}
\label{eqs:part}
\begin{align}
  s\vec{R}'(s)-\vec{r}'(0) &= A'\vec{R}'(s)+\vec{a} R_d(s)+\vec{c}'/s  \\
  s R_d(s)-r_d(0)          &= \vec{b}^T\vec{R}'(s)+a_{dd}R_d(s)+c_d/s.
\end{align}
\end{subequations}
If $s$ is not an eigenvalue of $A'$ then the first equation gives
\begin{equation}
  \vec{R}'(s) = (sI-A')^{-1} [\vec{r}'(0)+\vec{a} R_d(s)+\vec{c}'/s ]
\end{equation}
and inserting this into the second equation yields a non-linear equation
in $s$ that depends only on $R_d(s)$,
\begin{equation}
  0 = \vec{b}^T(sI-A')^{-1} [s\vec{r}'(0)+ s\vec{a} R_d(s)+\vec{c}']
       -s^2 R_d(s)+sr_d(0)+ s a_{dd} R_d(s)+c_d,
\end{equation}
which can be rewritten as
\begin{equation}
  \label{eq:main}
  R_d(s) = \frac{\vec{b}^T(sI-A')^{-1} [s\vec{r}'(0)+\vec{c}'] +sr_d(0)+c_d}
                {s^2 - s(\vec{b}^T(sI-A')^{-1} \vec{a} + a_{dd})}. 
\end{equation}
The equation must be satisfied for all $s$.  $R_d(s)$ is a rational
function and we can therefore expand the numerator and denominator in
terms of powers of $s$ and match the coefficients, leading to a set of
algebraic equations for the coefficients.  We can identify which
parameters of $A$ contribute to the RHS of Eq.(\ref{eq:main}), hence
determine what information can be extracted in principle.

In practice we can use knowledge about $A$ to infer the structure of the
expected signal.  From Sec.~\ref{sec2} we know that in the generic case
$A$ has $N^2-1$ distinct eigenvalues.  More precisely, we expect $N-1$
real eigenvalues and $(N^2-N-2)/2 = N(N-1)/2-1=:p$ pairs of complex
conjugate eigenvalues and thus a signal $\vec{r}(t)$ of the form
\begin{equation}
  r_d(t) = c_0 + \sum_{n=1}^{N-1} c_n e^{\gamma_n' t}
               + \sum_{n=1}^{p} e^{\gamma_n t} [a_n\cos(\omega_n t) + b_n \sin(\omega_n t)].
\end{equation}
Given stroboscopic measurement data $r_d(t_k)$ we can now use parameter
estimation techniques to estimate $\omega_n$, $\gamma_n$ and $\gamma_n'$ 
and the coefficients $a_n$, $b_n$ and $c_n$.

The approach above can be generalised to arbitrary partitioning of
the observed signals and in principle we could consider more general
observables.

\section{Application: Identification of Qubit Dynamics}

We illustrate the ideas above by applying them to a qubit evolving under
general Markovian dynamics. Taking
\begin{equation}
 \sx = \begin{pmatrix} 0 & 1 \\ 1 & 0 \end{pmatrix}, \quad
 \sy = \begin{pmatrix} 0 & -i \\ i & 0 \end{pmatrix}, \quad
 \sz = \begin{pmatrix} 1 & 0 \\ 0 & -1 \end{pmatrix}
\end{equation}
to be the unnormalised Pauli matrices, we can expand the Hamiltonian
$H=h_0I_2+h_x\sx+h_y\sy+h_z\sz$, where $h_j(j=0,x,y,z)$ are real.
Choosing $F_n$ to be dissipation operators to be the normalised Pauli
matrices, $F_1=\frac{1}{\sqrt{2}}\sigma_x$,
$F_2=\frac{1}{\sqrt{2}}\sigma_y$, $F_3=\frac{1}{\sqrt{2}}\sigma_z$,
$F_4=\frac{1}{\sqrt{2}}I_2$, and setting $f_{mn}=f^R_{mn}+if^I_{mn}$,
where $(f^R_{mn})$ is a real symmetric matrix and $(f^I_{mn})$ are
real-anti-symmetric matrix, the general Bloch operators take the form
\begin{equation}
 A= \begin{pmatrix}
-(f^R_{22}+f^R_{33})& f^R_{12}-2h_z        & f^R_{13}+2h_y\\
  f^R_{12}+2h_z     & -(f^R_{11}+f^R_{33}) & f^R_{23}-2h_x\\
  f^R_{13}-2h_y     & f^R_{23}+2h_x        & -(f^R_{11}+f^R_{22})
\end{pmatrix}, \quad
 \vec{c} = \sqrt{2}\begin{pmatrix} f^I_{23} \\ -f^I_{13} \\ f^I_{12} \end{pmatrix}.
\end{equation}
The anti-symmetric part of $A$ determines the Hamiltonian while the
symmetric part of $A$ determines the real parts of the dissipation
coefficients $f_{mn}$ and the inhomogeneous part determines the imaginary
part of $f_{mn}$.

Generically, we expect one real eigenvalue $-\delta$ and a pair of
complex eigenvalues $-\gamma\pm i\omega$ with $\delta,\gamma\ge0$.
Therefore we choose the basis functions for the Bayesian estimation
\begin{align}
  \label{eq:basis2}
   g_0(t) = 1, \quad
   g_1(t) = e^{-\gamma t} \cos(\omega t), \quad
   g_2(t) = e^{-\gamma t} \sin(\omega t), \quad
   g_3(t) = e^{-\delta t}, \quad
\end{align}
i.e., in particular the signal should be a linear combination of these
basis functions $s(t)=a_0 g_0(t) + a_1 g_1(t) + a_2 g_2(t) + a_3 g_3(t)$, 
and the Laplace transform of the signal should be the corresponding
linear combination of
\begin{align*}
   G_0(s) = \frac{1}{s}, \quad
   G_1(s) = \frac{s+\gamma}{(s+\gamma)^2+\omega^2}, \quad
   G_2(s) = \frac{\omega}{(s+\gamma)^2+\omega^2}, \quad
   G_3(s) = \frac{1}{s+\delta}.
\end{align*}
The parameter estimation in all of the following cases uses a standard
Bayesian estimation technique which involves orthogonal projections of
the data onto the basis functions and optimisation to maximise the
log-likelihood, this method being described more fully
elsewhere~\cite{TwoQubits,QutritDecoh}.

\subsection{Identification given full Bloch vector information}

\begin{figure}
\center\includegraphics[width=0.8\textwidth]{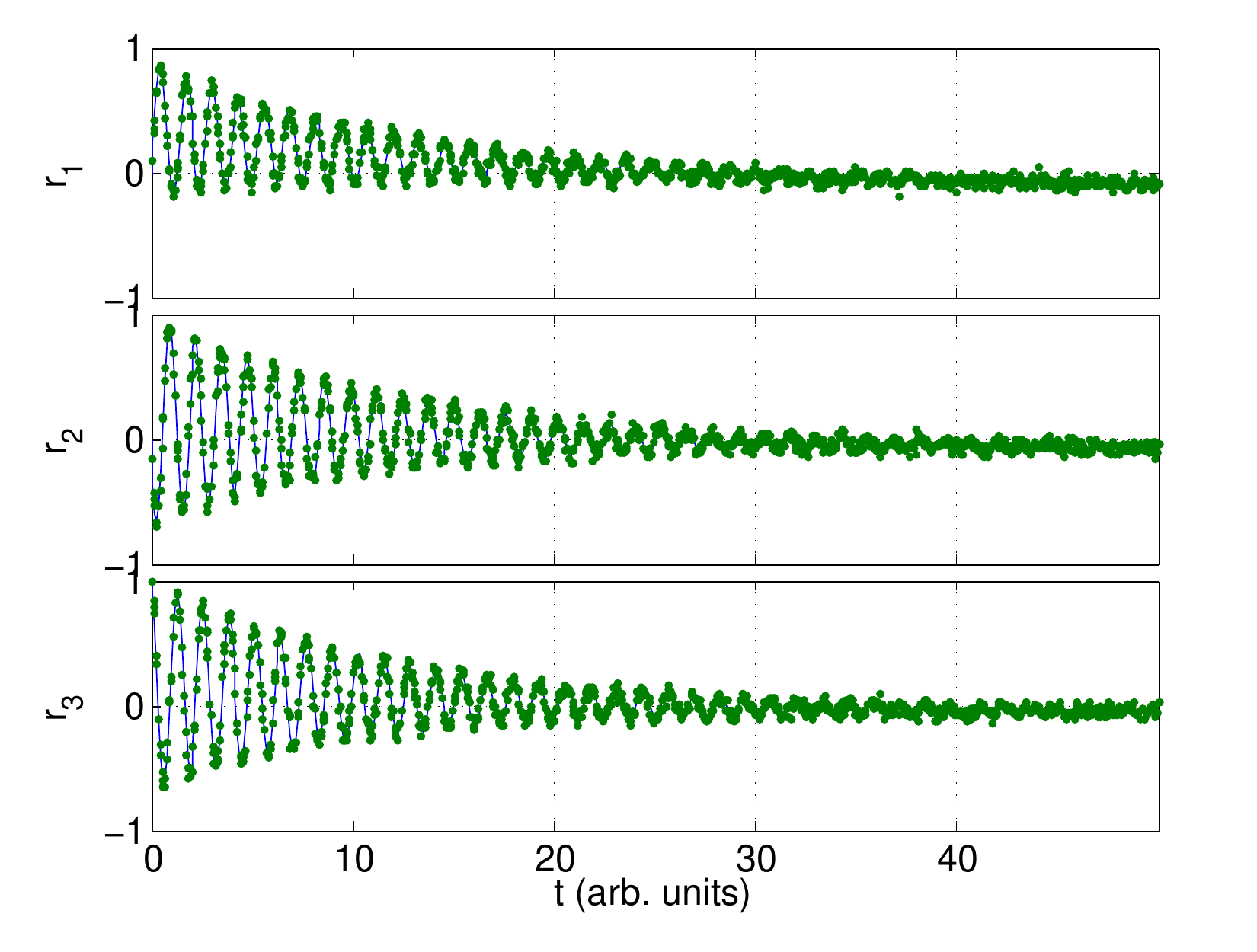}
\caption{Model 1, $T=50$, $N=1000$ and $N_e=1000$. Single input state
  but full tomographic data. For longer sampling times, the signal to
  noise ratio is too low and reconstruction accuracy suffers.}
\label{fig:traces1}
\end{figure}

We first consider a generic qubit system for which the Bloch matrix $A$
has three distinct eigenvalues.  The system is prepared in a fixed
initial state $\vec{r}_0$ and allowed to evolve under the unknown
dynamics. After evolving for different times, $t_k$, the components of
the Bloch vector $r_1(t_k)=\ave{\sx(t_k)}$, $r_2(t_k)=\ave{\sy(t_k)}$,
$r_3(t_k)=\ave{\sz(t_k)}$ are measured. For each time point the
experiment is repeated $N_e$ times to determine the relative frequency
of measurement outcomes $0$ and $1$. If we can measure all three
components of the Bloch vector (equivalent to QST of a qubit) then we
obtain three data traces of a form shown in Figure~\ref{fig:traces1}.

\begin{figure}
\center\includegraphics[width=\textwidth]{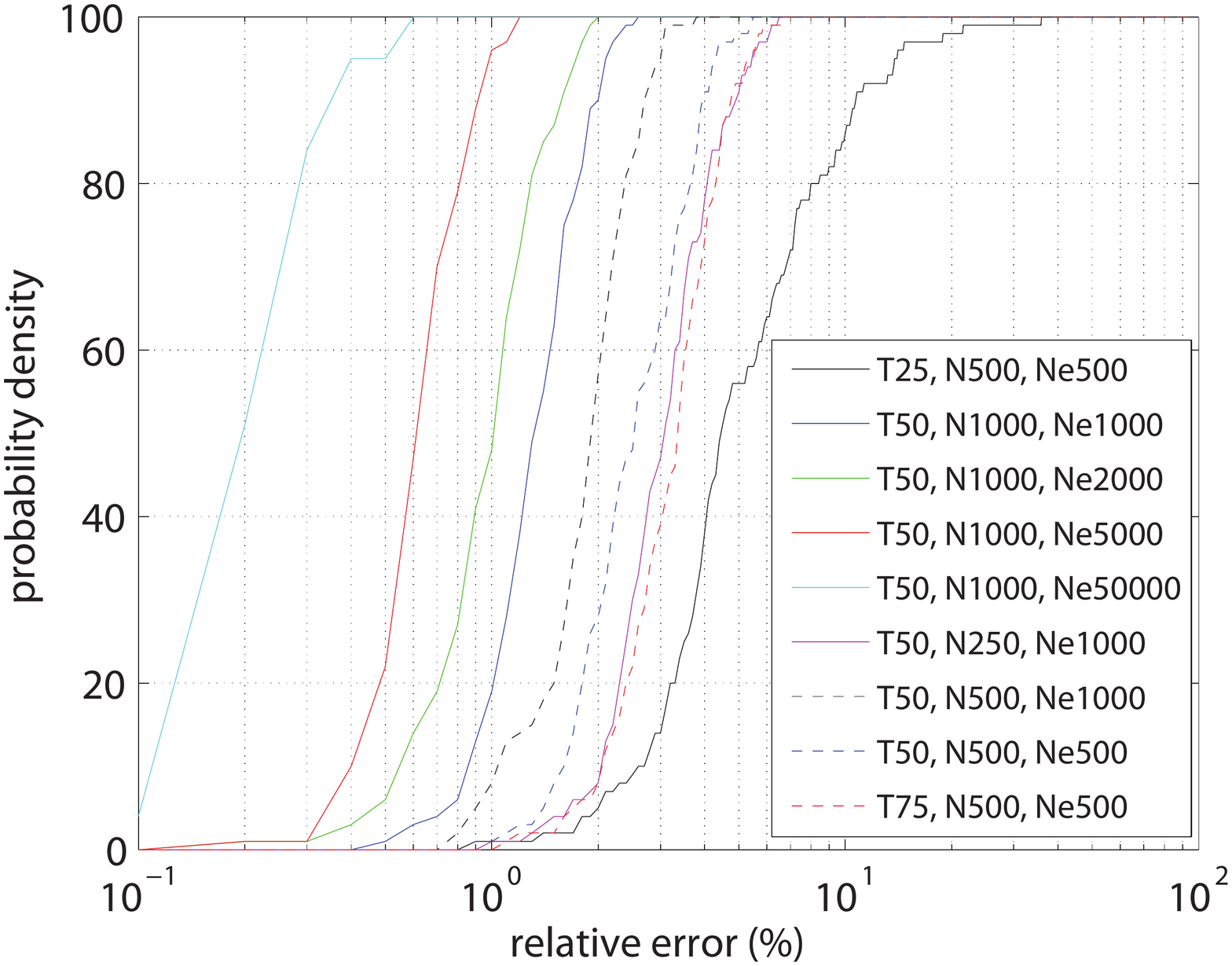}
\caption{Reconstruction-error cumulative probability density for
model~1. We examined different sampling times, sampling points and
experiment repetitions. As expected, as the noise of each data point
decreases, and the number of time samples increases, the reconstruction
accuracy diminishes. Different system models give similar error
probability plots.}  \label{fig:probDensity1}
\end{figure}

From these data traces, we then estimate the underlying signal
parameters and express the signal in the form~(\ref{eq:r(t)}) and
consequently reconstruct the matrix A and $\vec{c}$ using the technique
outlined in Section~\ref{sec2}. Figure~\ref{fig:probDensity1} shows the
cumulative probability density of the relative root-mean-squared error
of the reconstructed Bloch matrix $A$ and constant $\vec{c}$ for a
typical generic system, specifically (Model 1)
\begin{equation*}
 \label{eq:model1}
  A = \begin{pmatrix}
   -0.0650  &  -2.0000 &   2.0300\\
    2.0000  &  -0.0650 &  -4.0000\\
   -1.9700  &   4.0000 &  -0.0900
 \end{pmatrix}, \quad
 \vec{c} = \begin{pmatrix} -0.0424 \\ 0 \\ 0.0636 \end{pmatrix}, \quad
 \vec{r}_0 = \begin{pmatrix} 0 \\ 0 \\ 1 \end{pmatrix}
\end{equation*}
for different choices of the signal length, number of samples $N$ and
experiment repetitions $N_e$. A set of 100 randomly generated data
traces were used for the reconstruction.  Time samples for each run are
chosen by low-discrepancy sampling~\cite{LDS}.  As the figure shows the
median error for $N=N_e=1000$ is $1.5\%$; for $N=1000$, $N_e=5000$ the
relative error of more than $95\%$ of all runs is below $1\%$.  There is
an optimal range of the signal length around $T=50$ for this system. The
reconstruction error increases for substantially shorter or longer
signals; in the latter case this is due to the signal vanishing.
Comparing the results for $T=50$ and $N=2000$, $N_e=1000$ and $N=1000$,
$N_e=2000$ shows that there is virtually no difference in the
probability density, suggesting that it makes little difference if we
increase the number of time samples or the signal to noise ratio of the
sample points.

\subsection{Identification based on single component of Bloch vector}

Suppose we can only measure one component of the Bloch vector in a fixed
measurement basis.  This is typically the case where there is a
physically preferred or engineered mechanism and before coherent
rotations of the state are possible. Without loss of generality assume
we measure $r_d(t)=r_z(t)$, for instance by atomic population
measurement by fluorescence shelving~\cite{OxfordIon2008}.  We can write
the Laplace transform of the signal as
\begin{equation}
\label{eq:Rd}
  R_d(s) 
  = \frac{a_0}{s} + \frac{a_1(s+\gamma) + a_2\omega}{(s+\gamma)^2+\omega^2} 
                  + \frac{a_3}{s+\delta} 
  = \frac{C_3 s^3 + C_2 s^2 + C_1 s + C_0}{s^4 + D_3 s^3 + D_2 s^2 + D_1 s},
\end{equation}
Assuming we have estimated the coefficients $a_0$, $a_1$, $a_2$, $a_3$
as well as $\omega$, $\gamma$ and $\delta$ using signal parameter estimation,
the constants $C_k$, $D_k$ for $k=0,1,2$ are determined from the
observed signal
\begin{subequations}
\label{eqs:coeff1}
\begin{align}
  C_0 &= a_0 \delta (\gamma^2 + \omega^2) \\
  C_1 &= a_0 (\gamma^2+2\gamma\delta+\omega^2) 
       + a_1 \delta\gamma + a_2 \delta\omega + a_3 (\gamma^2+\omega^2) \\
  C_2 &= a_0(\delta +2\gamma) + a_1(\gamma+\delta) + a_2\omega + a_3 2\gamma\\
  C_3 &= a_0+a_1+a_3\\
  D_1 &= \delta(\gamma^2+\omega^2) \\
  D_2 &= 2\gamma\delta + \gamma^2 +\omega^2 \\
  D_3 &= 2\gamma +\delta 
\end{align}
\end{subequations}
Comparing with (\ref{eq:main}) for a generic single-qubit Bloch matrix
$A=(a_{mn})$ shows that the constants must satisfy 
\begin{subequations}
\label{eqs:coeff2}
\begin{align}
C_0 =& (a_{21} a_{32}-a_{22}a_{31})c_1
       + (a_{12}a_{31}-a_{11}a_{32})c_2 + (a_{11}a_{22}-a_{12}a_{21})c_3\\
C_1 =& a_{31} c_1 + a_{32} c_2 -(a_{11}+a_{22}) c_3
       + (a_{32}a_{21}-a_{31}a_{22}) r_x(0) \nonumber\\
     & + (a_{31}a_{12}-a_{32}a_{11}) r_y(0)
       + (a_{11}a_{22}-a_{12}a_{21}) r_z(0)\\
C_2 =& c_3 + a_{31} r_x(0) + a_{32} r_y(0)-(a_{11}+a_{22}) r_z(0) \\
D_1 =&  a_{13}a_{31}a_{22} - a_{12}a_{23}a_{31} - a_{13}a_{21}a_{32}
      + a_{11}a_{23}a_{32} + a_{12}a_{21}a_{33} - a_{11}a_{22}a_{33}\\
D_2 =& -a_{12}a_{21} - a_{13} a_{31} - a_{23}a_{32} + a_{22} a_{33} + a_{11}(a_{22}+a_{33})\\
D_3 =& -a_{11} -a_{22} -a_{33}
\end{align}
\end{subequations}
As the constants $C_k$, $D_k$ and the initial state $r_x(0)$, $r_y(0)$
and $r_d(0)$ are known, we have a system of six non-linear algebraic
equations for the coefficients of $A$ and $\vec{c}$.  Given that $A$ in
general has $9$ coefficients and $\vec{c}$ has three, these equations do
not determine $A$ uniquely, although we may be able to effectively
identify all unknown parameters if we have prior information about the
dominant types of dissipation, effectively restricting the form of $A$
and $\vec{c}$. This may be derived from physically motivated grounds or
from prior measurements of similar systems. We apply this to some
examples below.

\subsubsection{Dephasing in a general basis}
\label{sec:D1}

\begin{figure}
\center\includegraphics[width=0.8\textwidth]
{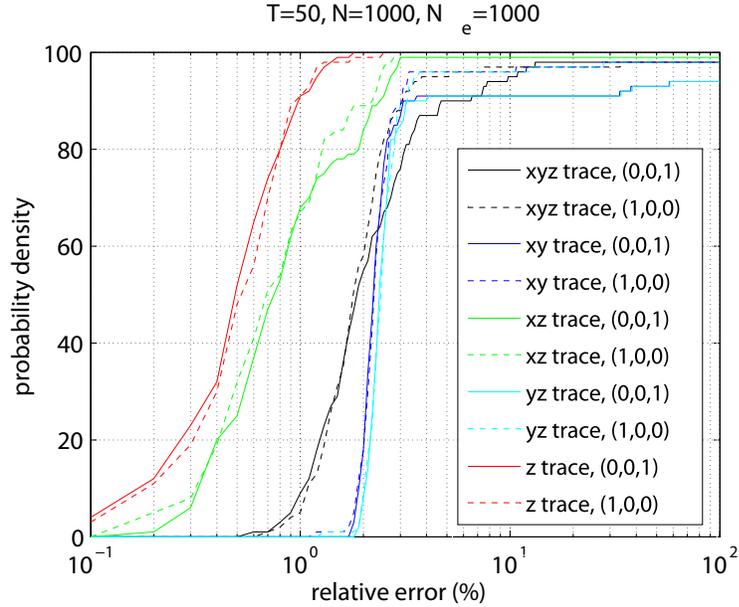}
\caption{Comparison of cumulative probability density for the relative
  Hilbert-Schmidt error of estimated Bloch matrix $A$ for Model 2. We
  examined different starting states and reconstructions based on full
  or partial observable information. The reconstruction based on full
  XYZ information did not assume any particular decoherence model,
  unlike the reconstruction based on only partial observables. Initial
  states were either $\vec{r}_0=(1,0,0)^T$ or
  $\vec{r}_0=(0,0,1)^T$. Partial observation was either the $z$ trace
  only, $y$ and $z$, $x$ and $z$, or $x$ and $y$ traces.}
\label{fig:errorDensity1}
\end{figure}


Consider the special case of a qubit subject to a fixed but unknown
Hamiltonian $H=\tfrac{1}{2}(h_x\sx+h_y\sy+h_z\sz)$ and dephasing
operator $V=\tfrac{1}{\sqrt{2}}(\alpha\sx + \beta\sy + \gamma \sz)$,
which corresponds to dephasing in an arbitrary basis. In this case
the master equation can be written in Lindblad form
\begin{equation}
   \label{eq:LME2} 
  \tfrac{d}{dt}\rho(t)
  = -\tfrac{i}{\hbar}[H,\rho] + V\rho(t)V^\dag - \tfrac{1}{2}\{V^\dag V,\rho(t)\}
\end{equation}
and expanding with respect to the normalised Pauli matrices, the
corresponding terms in the Bloch equation are
\begin{equation}
 A = \begin{pmatrix}
   -(\beta^2+\gamma^2) & -h_z+\alpha\beta     &  h_y+\alpha\gamma \\
   h_z+\alpha\beta     & -(\alpha^2+\gamma^2) & -h_x+\beta\gamma \\
  -h_y+\alpha\gamma    & h_x+\beta\gamma      & -(\alpha^2+\beta^2)
 \end{pmatrix}, \quad
 \vec{c}=\vec{0}.
\end{equation}
i.e., we have a total of six parameters.  We see from (\ref{eqs:coeff2})
that for $\vec{c}=0$ there are only five non-trivial signal parameters.
Thus, a single trace is not sufficient to completely determine all of
the model parameters.  If the system is initialised in a measurement
basis state $\vec{r}_0=(0,0,1)^T$ then (\ref{eqs:coeff1}) and
(\ref{eqs:coeff2}) give explicitly
\begin{subequations}
\label{eqs:coeff3}
\begin{align}
C_0 =& 0\\
C_1 =& h_z^2 + \gamma^2(\alpha^2+\beta^2+\gamma^2)\\
C_2 =& \alpha^2+\beta^2+2\gamma^2 \\
D_1 =& h_z^2 (\alpha^2+\beta^2) + h_y^2 (\alpha^2+\gamma^2) + h_x^2(\beta^2+\gamma^2)\nonumber\\
     &-2 h_x h_z \alpha\gamma - 2 h_x h_y \alpha\beta - 2 h_z h_y \beta\gamma \\
D_2 =& h_x^2 + h_y^2 + h_z^2 + (\alpha^2+\beta^2)^2 + \gamma^2(2\alpha^2+2\beta^2+\gamma^2)\\
D_3 =& 2(\alpha^2+\beta^2+\gamma^2) = 2(C_2-\gamma^2)
\end{align}
\end{subequations}
We cannot determine all 6 model parameters, the observed signal is
invariant under a rotation around the z-axis, hence we cannot determine
the azimuthal components of the Hamiltonian and dephasing
simultaneously, only their relative orientation.  This is similar to
the original situation of Hamiltonian identification where there is a
``gauge symmetry'' which exists due to only being able to initialise and
measure in a fixed basis, i.e. we do not have a phase reference to
measure the $x$ and $y$ components of the dynamics separately. However, we
can determine $h_x^2+h_y^2$ and $\alpha^2+\beta^2$. We can ``gauge
fix'', e.g. $h_y=0$, and define $\alpha$ and $\beta$ in reference to the
projection of the Hamiltonian onto the $x-y$ plane.

We have solved for the analytic solutions of (\ref{eqs:coeff3}) but
due to signal noise, these may not be exactly consistent. We therefore
implemented the non-linear system of equations by minimising the
least-squares errors using numerical optimisation starting with an
initial guess derived from the analytic solutions.
We plot the cumulative probability distribution for the errors in
Figure~\ref{fig:errorDensity1} for Model 2,
\begin{equation*}
       A = \begin{pmatrix}
	    -0.08 & -1.98 &  0.02\\
            2.02  & -0.05 & -1.96\\
            0.02  & 2.04  & -0.05
	   \end{pmatrix}, \quad
 \vec{c} = \begin{pmatrix} 0 \\ 0 \\ 0 \end{pmatrix}, 
\end{equation*}
which corresponds to $h_x=h_z=2$, $h_y=0$ and $\alpha=0.1$,
$\beta=\gamma=0.2$. We implemented various initial states, observables and prior knowledge.



\subsubsection{Dephasing and Relaxation in Measurement basis}
\label{sec:R1}

\begin{figure}
\center\includegraphics[width=0.8\textwidth]
{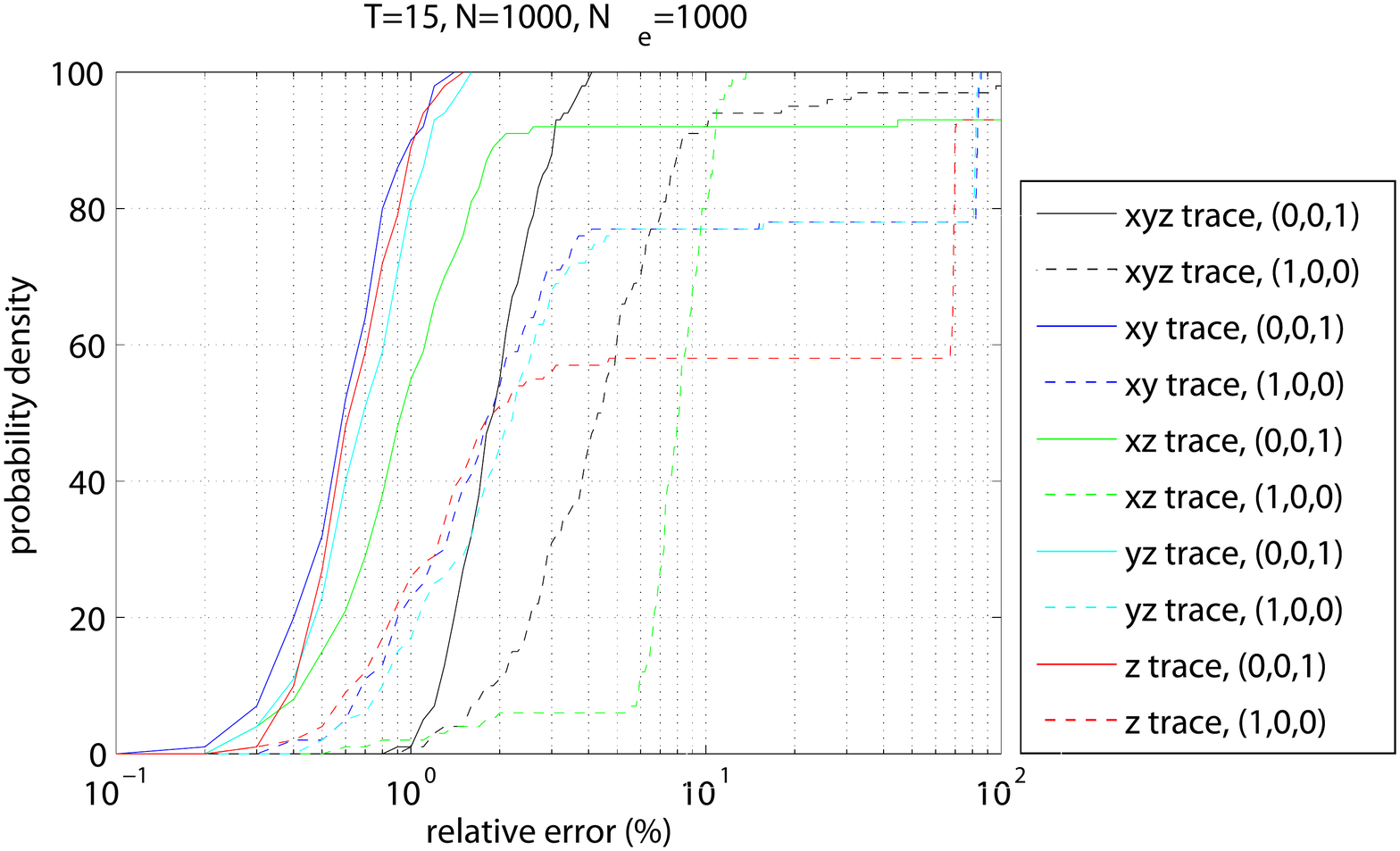}
\caption{Comparison of probability density for the relative
  Hilbert-Schmidt norm error of estimated Bloch matrix $A$ for Model
  3. Reconstruction procedures as in
  Figure~\ref{fig:errorDensity1}. The optimum sampling time is shorter
  than for Model 2 as the signal decays faster.}
\label{fig:errorDensity2}
\end{figure}

\begin{figure}
\center\includegraphics[width=0.8\textwidth]
{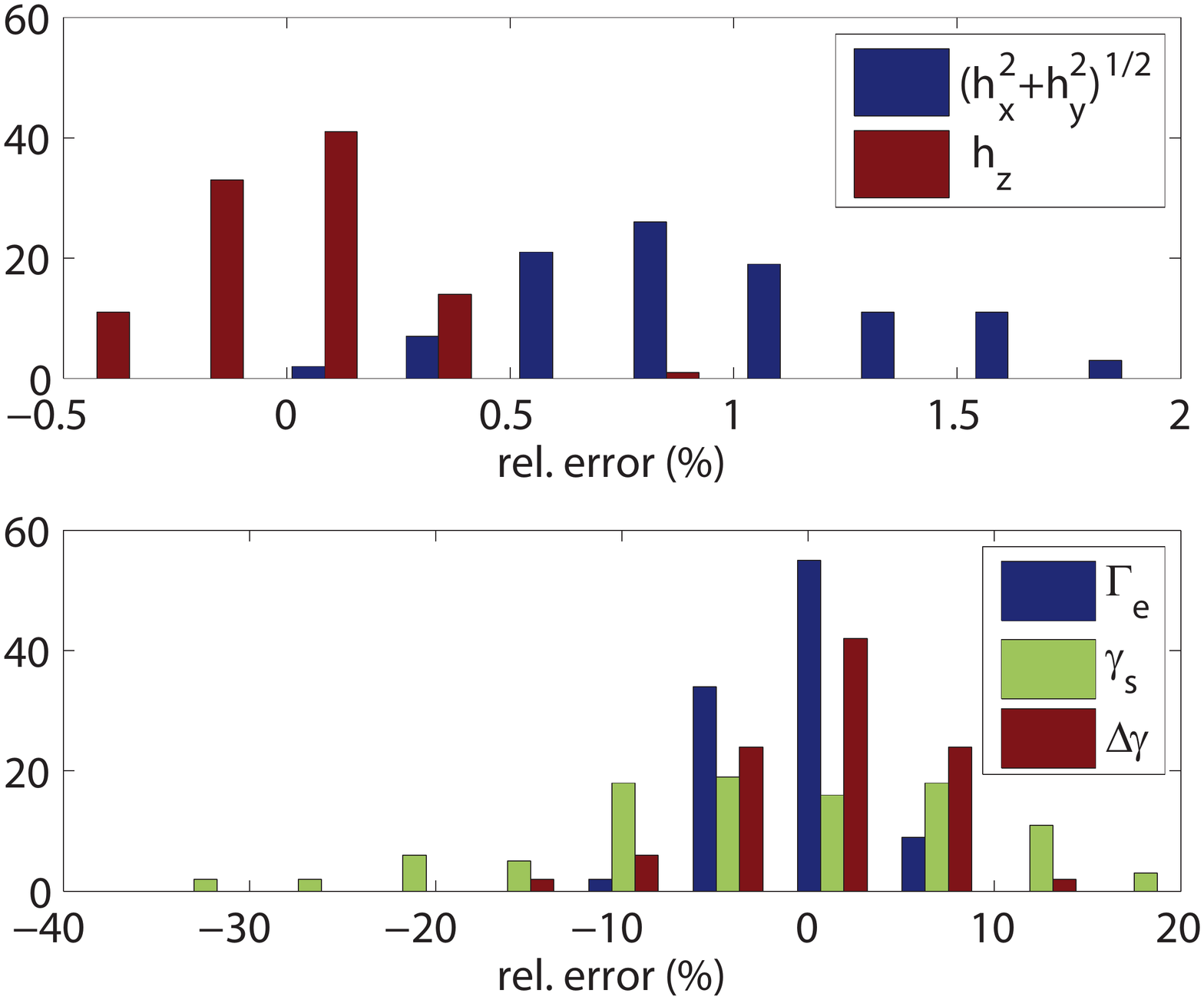}
\caption{Error histogram plot (Model 3, $r_0=(0,0,1)^T$, $z$-trace,
  $N=N_e=1000$, $T=15$). The absolute errors in the elements of matrix
  A were comparable with each other, but as the Hamiltonian parameters
  are much larger than the decoherence parameters, the relative errors
  of the latter are commensurately greater. Due to rotational symmetry
  around z, we can only determine $\sqrt{h_x^2+h_y^2}$.}
\label{fig:error_hist2}
\end{figure}

We also consider the special case of a qubit subject to a fixed
but unknown Hamiltonian $H=\tfrac{1}{2}(h_x\sx+h_y\sy+h_z\sz)$, and
relaxation and decoherence along the measurement basis governed by the
Lindblad equation
\begin{equation}
   \label{eq:LME3} 
  \frac{d}{dt}\rho(t) = -\frac{i}{\hbar}[H,\rho] 
                        + \sum_{k=1}^3 V_k\rho(t)V_k^\dag 
                              - \tfrac{1}{2}\{V_k^\dag V_k,\rho(t)\}
\end{equation}
with dephasing operator $V_1=\sqrt{\Gamma}\sz$, and relaxation
operators $V_2=\sqrt{\gamma_1}\ket{0}\bra{1}$ and
$V_3=\sqrt{\gamma_2}\ket{1}\bra{0}$ where we include both spontaneous
emission and absorption.  Again, expanding with respect to the
normalised Pauli basis the Bloch operators take the form
\begin{equation}
 A = \begin{pmatrix}
   -\Gamma_{\rm eff} &  -h_z &  h_y\\
   h_z  & -\Gamma_{\rm eff}  & -h_x\\
  -h_y  & h_x   & -\gamma_s
 \end{pmatrix}, \quad
 \vec{c}= \begin{pmatrix} 0 \\ 0 \\ \Delta\gamma \end{pmatrix}
\end{equation}
where $\Gamma_{\rm eff} = -2\Gamma - \frac{1}{2}\Delta\gamma$,
$\Delta\gamma = \tfrac{1}{\sqrt{2}}(\gamma_1-\gamma_2)$, $\gamma_s =
\gamma_1 + \gamma_2$.  For $\vec{r}_0=(0,0,1)^T$ the equations are
\begin{subequations}
\label{eqs:coeff1a}
\begin{align}
C_0 =& (\Gamma_{\rm eff}^2 + h_z^2)\Delta\gamma\\
C_1 =& 2 \Gamma_{\rm eff} \Delta \gamma +\Gamma_{\rm eff}^2+h_z^2\\
C_2 =& \Delta\gamma + 2\Gamma_{\rm eff} \\
D_1 =& \gamma_s (h_z^2+\Gamma_{\rm eff}^2) + (h_x^2+h_y^2)\Gamma_{\rm eff}\\
D_2 =& \Gamma_{\rm eff}^2 - 2\Gamma_{\rm eff}\gamma_s + h_x^2 + h_y^2 + h_z^2\\
D_3 =& 2\Gamma_{\rm eff}+\gamma_s.
\end{align}
\end{subequations}
Analytical solutions to these have also been found but as in the
previous case, numerical optimisation is required due to signal
noise. Again, we can only determine $h_x^2+h_y^2$ and not $h_x$ and
$h_y$ individually if we start off with $\vec{r}_0=(0,0,1)^T$ and
measure $r_z$. If we start off with $\vec{r}_0=(1,0,0)^T$, we obtain the
same expressions as in~(\ref{eqs:coeff1a}) except for the two equations
\begin{subequations}
\label{eqs:coeff1b}
\begin{align}
C_2 =& \Delta\gamma - h_y \\
C_1 =& 2\Gamma_{\rm eff} \Delta\gamma +  h_x h_z -  h_y\Gamma_{\rm eff},
\end{align}
\end{subequations}
which allow us in principle to identify all six parameters.  However,
the equations appear to be sensitive to noise and often lead to
incorrect model parameters.  This behaviour is a subject for further
research.  The results for Model 3
\begin{equation*}
       A = \begin{pmatrix}
	     -0.25  & -3.00 &  2.0  \\
              3.00  & -0.25 & -1.0  \\
             -2.00  & -1.00 &  0.1
	   \end{pmatrix}, \quad
 \vec{c} = \begin{pmatrix} 0 \\ 0 \\ 0.1/\sqrt{2} \end{pmatrix}, 
\end{equation*}
are shown in Figure~\ref{fig:errorDensity2} for various initial
states, observables, and prior knowledge. Figure~\ref{fig:error_hist2}
shows the histogram of the error distribution of the model parameters
for one reconstruction setting.



\subsection{Identification given two components of the Bloch vector}

In some systems, e.g. bulk nuclear magnetic resonance, determination
of the $x$ and $y$ components of the Bloch vector is easy to measure.
In this case, we assume we can measure both $r_x(t)$ and $r_y(t)$.
Inserting (18b) into (18a) gives
\begin{equation*}
  s\vec{R}'(s)-\vec{r}'(0) 
 = A'\vec{R}'(s)+\vec{a}[\vec{b}^T\vec{R}'(s)+a_{dd}R_d(s)+c_d/s + r_d(0)]
+\vec{c}'/s.
\end{equation*}
Solving for $\vec{R}'(s)$ gives
\begin{equation}
\vec{R}'(s)  =
[sI'-A'-\vec{a}\vec{b}^T/(s-a_{dd})]^{-1}
(\vec{r}'(0)+\vec{c}'/s + \vec{a}(c_d/s+r_d(0))/(s-a_{dd})
\end{equation}
from which we can obtain two explicit rational equations for $R_x(s)$
and $R_y(s)$.  The denominators of both are the same and the
coefficients of the polynomials on the top and bottom can be related to
the observed signals as before.

For the two examples in sections~\ref{sec:D1} and~\ref{sec:R1}, we
implemented the reconstruction procedure, the results of which are
presented in Figures~\ref{fig:errorDensity1}
and~\ref{fig:errorDensity2}.  For the sake of comparison we considered
not only $x$ and $y$ but different combination of two traces including
$y$ and $z$, and $x$ and $z$ observable information for the
reconstruction.  The general observation is that for the systems
considered the $z$-trace appeared to be the best choice for the
reconstruction.  The addition of the extra information in the form of
the $x$ or $y$ trace did not improve the reconstruction accuracy over
reconstruction using the $z$ trace alone for these examples, and using
the $x$ and $y$ traces instead appear to result in substantially worse
reconstruction results.  There are also initial state effects with 
initialisation in the $z$-eigenstate being decided preferable compared
to initialisation in the $x$-eigenstate in the second case, for example.

\section{Summary and Discussion}

We have studied the problem of identification of the dynamics of open
systems in a Markovian environment without recourse to quantum process
tomography.  By utilising stroboscopic measurements over time we showed
that the system dynamics is completely determined by the evolution of a
single initial state, at least in the generic case when $A$ has distinct
eigenvalues and the initial state has overlap with all eigenspaces of
$A$, i.e., we can in principle fully determine the Bloch operator $A$
and inhomogeneous term $\vec{c}$ from stroboscopic measurements of the
state at different times.  Analysis of non-generic cases suggests that
this result basically still holds in most cases and we can give specific
conditions under which full information about the dynamics can be
obtained.  For systems governed by Markovian evolution this approach is
preferable to process tomography for several reasons.  First, it does
not require initialisation in a large set of different basis states.
Second, it avoids the rather messy procedure of extracting the dynamical
generators from process tomography data, which is susceptible noise.
Third, it is more efficient than performing process tomography at
numerous different times.

By going to the Laplacian domain, we can identify algebraic equations
relating the observed signals to the models parameters. In the absence
of a mechanism of directly sampling in the Laplacian domain, we can use
Bayesian signal estimation in the time domain to find a functional
representation of the measurements more amenable to Laplacian domain
analysis. The most important advantage of this approach is that it shows
that in many cases most or all model parameters can be obtained from far
less data, i.e., we do not even require full state tomography.  Rather
we can extract the same information from measurements of a smaller set
of observables.  This approach is especially useful if there is physical
information restricting the type of decoherence model we expect, and the 
dynamical generators depend on fewer parameters.  Our analysis of the
qubit cases shows that not only is it possible to extract all of the
system parameters from this restricted data but often this even leads
to better reconstruction results than using full state reconstruction.

In part these results can be explained by overfitting, i.e., trying to
fit the most general model given noisy data when the actual dynamics is
determined by a smaller set of parameters leads to fitting the noise and
results in worse model over all.  However, overfitting does not appear
to explain the peculiar dependence of the results on the choice of
initial state or observables noted earlier.  For many of the systems
considered the success of the reconstruction showed a strong dependence
on the choice of the initial state and the type of measurement, with
initialisation in a $z$-eigenstate of measurement of $\ave{\sigma_z}$
being distinctly preferable over other combinations such as measuring
$\ave{\sigma_x}$ or initialising in an $x$-eigenstate.  This effect was
most pronounced for system subject to relaxation in the measurement
basis.  In this case the choice of the initial state $(1,0,0)^T$ in
principle breaks the rotational symmetry, allowing the determination of
all six model parameters, instead of the five using the initial state
$(0,0,1)^T$.  However, in the former case the reconstruction often
fails, converging to a model which is distinctly different from the
actual model.  There appears to be a trade-off between the breadth of
the information one versus the quality of the information gathered but
the issue of the optimal choice of the initial state and observables
warrants further study, as do the choice of optimal signal lengths,
adaptive sampling, stability of reconstruction algorithms in the 
presence of noise among others.

\ack

SGS acknowledges financial support from EPSRC Advanced Research
Fellowship EP/D07192X/1.  DKLO acknowledges support from Quantum
Information Scotland network (QUISCO).  MZ is supported by the National
Science Foundation of China.  Part of this work was conducted while WZ
and EG were visiting SGS at Cambridge and we gratefully acknowledge NUDT
and the University of Cambridge for supporting the visit.

\appendix

\section{Non-diagonalisable Bloch matrices for $N=2$}
\label{App:non-generic}

As the complex eigenvalues of the Bloch matrix $A$ must occur in pairs
$\gamma\pm i\omega$, we can only have a non-trivial Jordan block for a
qubit system if $\omega=0$, i.e., the eigenvalues are real, and there
are five cases: $A=S J S^{-1}$ with
\begin{equation}
 J_1= \begin{pmatrix} \gamma_1 & 1 & 0 \\
                        0 & \gamma_1 &0 \\
                        0 & 0 & \gamma_2
      \end{pmatrix}, \quad
 J_2= \begin{pmatrix} \gamma_1 & 0 & 0 \\
                        0 & \gamma_1 &0 \\
                        0 & 0 & \gamma_2
      \end{pmatrix}, \quad
 J_3= \begin{pmatrix} \gamma & 1 & 0 \\
                        0 & \gamma & 1 \\
                        0 & 0 & \gamma
      \end{pmatrix}, \quad
 J_4= \begin{pmatrix} \gamma & 1 & 0 \\
                        0 & \gamma & 0 \\
                        0 & 0 & \gamma
      \end{pmatrix}, \quad
\end{equation}
and $J_5=\gamma I$.  The corresponding time evolution operators are
$e^{tA}=S e^{tJ} S^{-1}$ with
\begin{gather*}
  e^{tJ_1} = \begin{pmatrix} 
	      e^{t\gamma_1} & t e^{t\gamma_1} & 0 \\
              0 & e^{t\gamma_1} &0 \\
              0 & 0 & e^{t\gamma_2}
	     \end{pmatrix}, \quad
  e^{tJ_2} = \begin{pmatrix} 
	      e^{t\gamma_1} & 0 & 0 \\
              0 & e^{t\gamma_1} &0 \\
              0 & 0 & e^{t\gamma_2}
	     \end{pmatrix}, \\
  e^{tJ_3} = \begin{pmatrix} 
	      e^{t\gamma} & t e^{t\gamma} & \tfrac{t^2}{2} e^{t\gamma}\\
                        0 & e^{t\gamma} & t e^{t\gamma} \\
                        0 & 0 & e^{t\gamma}
	     \end{pmatrix}, \quad
  e^{tJ_4} = \begin{pmatrix} 
	      e^{t\gamma} & t e^{t\gamma} & 0\\
                        0 & e^{t\gamma} & 0 \\
                        0 & 0 & e^{t\gamma}
	     \end{pmatrix}, \quad
  e^{t J_5} = e^{\gamma t} I.
\end{gather*}

\textbf{Case 1.} There are two proper eigenvectors $\vec{x}_{11}$,
$\vec{x}_{2}$ and one generalized eigenvector $\vec{x}_{12}$ satisfying
$(A-\gamma_n I)\vec{x}_{n1}=0$ for $n=1,2$ and $(A-\gamma_1 I)^2
\vec{x}_{12}=0$.  Expanding with respect to these $\vec{\Delta}(0) =
\alpha_{11} \vec{x}_{11} + \alpha_{12} \vec{x}_{12} + \alpha_2
\vec{x}_2$ and using the Jordan form of $A$ above gives
\begin{align*}
  \vec{r}(t) &= \vec{r}_{\ss} + e^{t\gamma_1} (\alpha_{11}+\alpha_{12} t) \vec{x}_{11}
                     + e^{t\gamma_1} \alpha_{12} \vec{x}_{12}
                     + e^{t\gamma_2} \alpha_2 \vec{x}_2 \\
             &= \vec{r}_{\ss} 
                    + e^{t\gamma_1} t\alpha_{12} \vec{x}_{11} 
                    + e^{t\gamma_1} (\alpha_{11} \vec{x}_{11}+\alpha_{12} \vec{x}_{12})
                    + e^{t\gamma_2}\alpha_2 \vec{x}_{2}\\
             &= \vec{a}_0 + te^{t\gamma_1} \vec{a}_1
                    + e^{t\gamma_1} \vec{a}_2 + e^{t\gamma_2} \vec{a}_3
\end{align*}
Using parameter estimation we can in principle estimate $\gamma_n$,
$n=1,2$, as well as the coefficient vectors $\vec{a}_n$ for $n=0,1,2,3$.
$\vec{a}_0$ determines the steady state and $\vec{a}_1$ and $\vec{a}_3$
determine the two primary eigenvectors as we can absorb the factors
$\alpha_{12}$ and $\alpha_2$.  $\vec{a}_2$ is a generalised eigenvector
as $(A-\gamma_1 I) (\alpha_{11} \vec{x}_{11}+\alpha_{12}\vec{x}_{12})
=(A-\gamma_1 I) \alpha_{12}\vec{x}_{12}\neq 0$, $(A-\gamma_1 I)^2
(\alpha_{11} \vec{x}_{11}+\alpha_{12}\vec{x}_{12})=0$.  Thus we have
$S=(\vec{a}_1, \vec{a}_2, \vec{a}_3)$ and both eigenvalues and $A$ is
completely determined.

\textbf{Case 2.} There are two proper eigenvectors $\vec{x}_1$,
$\vec{x}_1'$ with eigenvalue $\gamma_1$ and one proper eigenvector
$\vec{x}_2$ with eigenvalue $\gamma_2$.  Expanding with respect to these
eigenvectors $\vec{\Delta}(0) = \alpha_1 \vec{x}_1 + \alpha_1' 
\vec{x}_1' + \alpha_2 \vec{x}_2$ and
\begin{align*}
  \vec{r}(t) &= \vec{r}_{\ss} + e^{t\gamma_1} (\alpha_1 \vec{x}_1+\alpha_1'\vec{x}_1')
                     + e^{t\gamma_2} \alpha_2 \vec{x}_2 \\
             &= \vec{a}_0 + e^{t\gamma_1} \vec{a}_1 + e^{t\gamma_2} \vec{a}_2
\end{align*}
Using parameter estimation we can estimate $\gamma_n$ as well as the
coefficient vectors $\vec{a}_n$ for $n=0,1,2$.  $\vec{a}_0$ determines
one steady state and $\vec{a}_1$ and $\vec{a}_2$ determine two of the
proper eigenvectors of $A$ but without further information we cannot
determine the third eigenvector.

\textbf{Case 3.}  In this case there is only one eigenvalue $\gamma$ and
one proper eigenvector $\vec{x}_{11}$ and two generalised eigenvectors
$\vec{x}_{12}$, $\vec{x}_{13}$.  Expanding with respect to these
generalised eigenvectors, $\vec{\Delta}(0) =\alpha_{11} \vec{x}_{11} +
\alpha_{12} \vec{x}_{12} + \alpha_{13} \vec{x}_{13}$, and using the
Jordan form gives
\begin{align*}
  \vec{r}(t) &= \vec{r}_{\ss} 
                 + e^{t\gamma} (\alpha_{11}+\alpha_{12} t + \tfrac{1}{2}\alpha_{13}t^2) \vec{x}_{11}
                 + e^{t\gamma} (\alpha_{12} + t \alpha_{13}) \vec{x}_{12}
                 + e^{t\gamma} \alpha_{13} \vec{x}_{13} \\
             &= \vec{r}_{\ss} + \tfrac{1}{2}\alpha_{13} t^2 e^{t\gamma} \vec{x}_{11}
                + t e^{t\gamma} (\alpha_{12} \vec{x}_{11} + \alpha_{13} \vec{x}_{12})
                + e^{t\gamma} (\alpha_{11} \vec{x}_{11}+\alpha_{12} \vec{x}_{12} + \alpha_{13} \vec{x}_{13})\\
             &= \vec{a}_{0} + t^2 e^{t\gamma} \vec{a}_1
                + t e^{t\gamma} \vec{a}_2 + e^{t\gamma} \vec{a}_3
\end{align*}
Using parameter estimation we can estimate $\gamma$ as well as the
coefficient vectors $\vec{a}_n$ for $n=0,1,2,3$.  Again $\vec{a}_0$
determines a steady
state. $\vec{a}_1=\tfrac{1}{2}\alpha_{13}\vec{x}_{11}$ is a proper
eigenvector of $A$, $\vec{a}_2$ and $\vec{a}_1$ are generalised
eigenvectors with $(A-\gamma I)^2 \vec{a}_2=0$ and $(A-\gamma I)^3
\vec{a}_3=0$, respectively.  $S=(\vec{a}_1, \vec{a}_2, \vec{a}_3)$ and
$\gamma$ and $A$ is completely determined.

\textbf{Case 4.}  In this case there is again only one eigenvalue
$\gamma$ but there are two proper eigenvectors $\vec{x}_1$, $\vec{x}_1'$
and one generalised eigenvector $\vec{x}_{12}$ with $(A-\gamma
I)\vec{x}_{12} \neq 0$ but $(A-\gamma I)^2 \vec{x}_{12}=0$. Expanding
with respect to these generalised eigenvectors, $\vec{\Delta}(0)
=\alpha_1 \vec{x}_1 + \alpha_1' \vec{x}_1' + \alpha_{12}\vec{x}_{1}$, 
and using the Jordan form gives
\begin{align*}
  \vec{r}(t) &= \vec{r}_{\ss} 
                + e^{t\gamma} [(\alpha_1+\alpha_{12} t) \vec{x}_1 
                + \alpha_{12} \vec{x}_{12} + \alpha_1' \vec{x}_1' ] \\
             &= \vec{r}_{\ss} 
                 + t e^{t\gamma} \alpha_{12} \vec{x}_1
                 + e^{t\gamma} (\alpha_1\vec{x}_1 + \alpha_1'\vec{x}_1' + \alpha_{12} \vec{x}_{12})\\
             &= \vec{a}_{0} + t e^{t\gamma} \vec{a}_1
                + e^{t\gamma} \vec{a}_2 
\end{align*}
Using parameter estimation we can estimate $\gamma$ as well as the
coefficient vectors $\vec{a}_n$ for $n=0,1,2$.  Again $\vec{a}_0$
determines a steady
state. $\vec{a}_1=\tfrac{1}{2}\alpha_{13}\vec{x}_{11}$ is a proper
eigenvector of $A$, $\vec{a}_2$ determines another generalised
eigenvector but the third eigenvector can again not be determined.

\textbf{Case 5.} In this case any vector is an eigenvector with
eigenvalue $\gamma$ and we have $\vec{r}(t) = \vec{r}_{\ss}+ e^{t\gamma}
\vec{\Delta}(0)$.  Using parameter estimation we can determine $\gamma$,
$\vec{a}_0$ and $\vec{\Delta}(0)$ and in this case this information is
sufficient to determine $A$ and $\vec{c}$ --- if it is known that we
have case 5.

Cases 1 and 3 can clearly be identified from the structure of the
observed signal.  A signal of the form observed in case 2 can also arise
if $A$ has three distinct (real) eigenvalues but the initial state is
such that $\vec{\Delta}(0)$ has no overlap with one of the eigenvectors.
Case 4 can also arise if $A$ has two real eigenvalues, one of which
corresponding to a Jordan block of size 2 but the initial state is such
that $\vec{\Delta}(0)$ has no overlap with one of the proper
eigenvectors.  Finally, a signal of the form observed in Case 5 can
arise not only when $A$ is a scalar matrix but also when
$\vec{\Delta}(0)$ does not have overlap with all (generalised)
eigenvectors of $A$.

\end{document}